\documentclass[12pt]{mn2e}
\usepackage[dvips]{graphicx}
\usepackage[dvips]{rotating}

\title[New gravitational lens ULAS~J082016.1+081216]
{A new gravitational lens from the MUSCLES survey: ULAS~J082016.1+081216}
\author[Jackson et al.]
{Neal Jackson$^{1}$, Eran O. Ofek$^{2}$, Masamune Oguri$^{3}$\\
$^{1}$ Jodrell Bank Centre for Astrophysics, University of Manchester,
Turing Building, Oxford Road, Manchester M13 9PL\\
$^{2}$Division of Physics, Mathematics and Astronomy, California Institute of
Technology, Pasadena, CA 91125, USA\\
$^{3}$Kavli Institute for Particle Astrophysics and Cosmology, Stanford
University, Menlo Park, CA 94025, USA\\
}
\input psfig.tex
\begin{document}
\maketitle
\begin{abstract}
We present observations of a new double-image gravitational lens system,
ULAS~J082016.1+081216, of image separation
2\farcs3 and high ($\sim$6) flux ratio. The system is selected from the
Sloan Digital Sky Survey spectroscopic quasar list using new high-quality
images from the UKIRT Deep Sky Survey (UKIDSS). The lensed quasar has a 
source redshift of 2.024, and we identify the lens galaxy as a faint red object
of redshift 0.803$\pm$0.001. Three other objects from the UKIDSS survey,
selected in the same way, were found not to be lens systems. Together 
with the earlier lens found using this method, the SDSS-UKIDSS lenses 
have the potential to significantly increase the number of quasar lenses
found in SDSS, to extend the survey to higher flux ratios
and lower separations, and to give greater completeness which is
important for statistical purposes. 
\end{abstract}


\large

\section{Introduction}

More than 100 cases of strong gravitational lensing are now known in which quasars are
multiply lensed by foreground galaxies, about the same quantity as the
number of galaxy-galaxy lensing systems. The two types of system have
different advantages. Systems with lensed galaxies are usually extended
and therefore typically provide more constraints on the first derivative
of the gravitational potential, as has been shown by the large survey of
such systems from the Sloan Digital Sky Survey, SLACS (Bolton et al.
2006; Koopmans et al. 2006; Bolton et al. 2008). On the other hand,
time delay measurements of variations in the images of lensed quasars 
provide a measurement of the combination of the Hubble constant $H_0$ 
(Refsdal 1964) and the average surface density of the lens in the annulus 
between the images used to determine the delay (Kochanek 2002).
Moreover, the selection effects are often different; galaxy-galaxy
systems such as the SLACS survey are usually selected based on the
lenses, whereas lensed quasars are usually selected based on the
sources. This has important implications for statistical studies.

In many cases, the statistics of a well-selected set of gravitational
lenses can provide important cosmological information. The original
application of source-selected lens samples, the determination of 
combinations of the cosmic matter density $\Omega_m$ and cosmological 
constant density $\Omega_{\Lambda}$ in units of the critical density 
(Fukugita et al. 1992, Maoz \& Rix 1993, Kochanek 1996) has now been largely superseded 
by other methods such as studies of the cosmic microwave background,
supernova brightness, and baryon acoustic oscillations. However, once
the global cosmological model is known, the statistics of gravitational
lensing can provide important information about the evolution of
galaxies. Early studies used the radio sample CLASS (Myers et al. 2003,
Browne et al. 2003) which contained 13 quasar lenses in a statistically
complete sample (22 lenses overall) of radio sources with 5-GHz flux
density $\geq$30~mJy. One major use of such samples is the
``lens--redshift'' test (Kochanek 1992) in which knowledge of the lens
and source redshifts and image separations can be used to make
inferences about galaxy evolution, given a global cosmology. 
This was used by Ofek, Rix \& Maoz (2003) and most recently Matsumoto 
\& Futamase (2008) to derive limits on the evolution of the galaxy number
density and velocity dispersion, in terms of the redshift evolution of 
a fiducial number density and velocity dispersion from a Schechter-like 
function. In surveys to date, the available sample of lenses is consistent
with no evolution up to $z\sim1$ and a standard
$\Lambda$CDM cosmology, but expansion of the sample is desirable in
order to enable a more stringent test. Capelo \& Natarajan (2007) study
the robustness of this test, concluding that larger and more
uniform samples of lenses, with complete redshift information and good
coverage of separation distributions, are required.

In recent years, larger samples have become available
by investigation of quasars from the Sloan Digital Sky Survey quasar
list (Schneider et al. 2007). These have been used by Inada and
collaborators (e.g. Inada et al. 2003; Inada et al. 2008) to discover 30 lensed
quasars to date, which form the SQLS (SDSS Quasar Lens Search, Oguri et
al. 2006). Optical
surveys are somewhat more difficult to carry out, in that the high
resolution needed to separate the components of the lens system is less
easily available in the optical; the CLASS survey, which had a limiting
lens separation of 0\farcs3, showed that the median lens separation is of the
order 0\farcs8. 

Although the SDSS covers a large fraction of the sky to a relatively
faint ($r\sim 22$) limiting magnitude, with the Legacy DR7 spectroscopy
now totalling 9380 square degrees, the PSF width of the images is
typically 1\farcs4. More recently the UKIRT Deep Sky Survey (UKIDSS,
Lawrence et al. 2007) has become available; the UKIDSS Large Area Survey
(ULAS) now covers just over 1000 square degrees to a depth of K=18.4
(corresponding to $R\sim24$ for a typical elliptical galaxy at $z=0.3$)
and, importantly, has a median seeing of 0\farcs8. UKIDSS uses the
UKIRT Wide Field Camera (WFCAM; Casali et al, 2007); the photometric
system is described in Hewett et al (2006), and the calibration is
described in  Hodgkin et al. (2009). The pipeline processing and science
archive are described in Irwin et al (2009, in prep) and Hambly et al
(2008).

We are therefore
conducting a programme (Major UKIDSS-SDSS Cosmic Lens Survey, or
MUSCLES) which aims to discover lenses difficult for or inaccessible to
the SQLS due to small separation, high flux ratio or a combination of
the two. We have used data from the UKIDSS 4th data release in this work.
In an earlier paper, we reported the discovery of the first
lens found in this way (ULAS~J234311.9$-$005034, Jackson, Ofek \& Oguri 2008).
Here we describe a second detection of a lens system, of relatively
large separation but with a relatively faint secondary. In section 2 we
describe the survey selection and observations. In section 3 we discuss
the results, including the three objects rejected as lenses and the
evidence that ULAS~J082016.1+081216 is a lens system. Finally, in
section 4 we revisit the survey selection in the light of the two lenses
discovered by the MUSCLES programme, to assess its potential to discover
new lenses which are of smaller separation and/or higher flux ratio.

\section{Sample selection and observations}

Objects were selected from the fourth Data Release (DR4) of UKIDSS, and
compared against the SDSS quasar catalogue (SDSS DR5, Schneider et al.
2007). Of the 77429 SDSS quasars, 6708 objects were identified, due mainly to the
limited area coverage of current UKIDSS. These were then inspected by
eye for extensions, although we are currently developing algorithms for
supplementing with objective selection from parameters fitted to the
UKIDSS images. We identified 150 good candidates, of which 14 had
already been ruled out by other observations (mainly SQLS), and seven
(not including ULAS~J234311.9$-$005034, Jackson 2008) 
were known lenses. The survey
rediscovered all known lenses in the current UKIDSS 
footprint\footnote{SDSS J080623.7+200632 (Inada et al. 2006), 
SDSS J083217.0+040405 (Oguri et al. 2008), 
SDSS J091127.6+055054 = RXJ0911+0551 (Bade et al. 1997), 
SDSS J092455.8+021925 (Inada et al. 2003), 
SDSS J122608.0$-$000602 (Inada et al. 2009, in prep), 
SDSS J132236.4+105239 (Oguri et al. 2008a), SDSS J135306.2+113805 (Inada et
al. 2006).}. Of the 129 remaining objects, one, ULAS~J234311.9$-$005034, 
was observed previously by us and found to be a lens (Jackson et al.
2008). In this work we describe observations of four further objects
from the candidate list.

These four objects were observed using the Keck-I telescope on Mauna Kea
on the night of 2009 February 17, using the LRIS-ADC double-beam imaging
spectrograph (Oke et al. 1995). They were selected as the most
convenient objects for observation at the available time, which appeared
on subjective examination to be the most likely lenses, and which had
estimated sizes which could be resolved by the seeing of the observations,
roughly 1$^{\prime\prime}$.  The blue 
arm of the spectrograph was used with a central wavelength of 430~nm, 
and the red arm with a central wavelength of 760~nm. A dichroic cutting 
between 560 and 570~nm was used to split the light between the two arms. 
A long slit of width 0\farcs7 was used, with
a position angle chosen so as to cover the extended structure seen in
the UKIDSS images. A list of objects observed together
with integration times is given in Table 1, and UKIDSS images of the
observed objects are presented in Fig. 1.

\begin{table*}
\begin{tabular}{cccccc} \hline
Object & $z_{SDSS}$ & $r_{SDSS}$ & Exp. (blue)/s & Exp. (red)/s & Separation/$^{\prime\prime}$\\ \hline
J033248.5$-$002155 & 1.713 & 18.36 & 1800 & 1650 & 1.1 \\
J034025.5$-$000820 & 0.619 & 20.13 & 1600 & 1560 & 1.4 \\
J082016.1+081216 & 2.024 & 18.97 & 1450 & 1400 & 1.9 \\
J091750.5+290137 & 1.816 & 18.07 & 1540& 1400 & 1.0 \\ \hline
\end{tabular}
\caption{Details of the Keck-I observations, showing the objects (with
names representing J2000 coordinates), the SDSS redshift and $r$ magnitude,
image separations (measured from the UKIDSS images) and the exposure times
in the blue and red arms. All observations were
carried out on the night of 2009 February 17 using the LRIS
spectrograph.}
\end{table*}

\begin{figure}
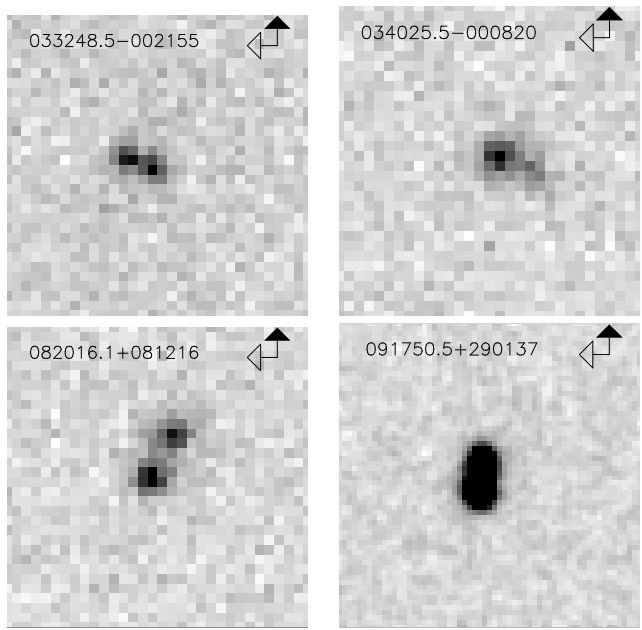

\begin{tabular}{cc}
\psfig{figure=033248.5-002155.ps,width=4cm,angle=-90}&
\psfig{figure=034025.5-000820.ps,width=4cm,angle=-90}\\
\psfig{figure=082016.1+081216.ps,width=4cm,angle=-90}&
\psfig{figure=091750.5+290137.ps,width=4cm,angle=-90}\\
\end{tabular}
\caption{UKIDSS images of the objects observed. Images are in the $H$-band
except for J091750.5+290137, which is in the $J$-band. All images have
North at the top and East on the left, and each image is 12\farcs8 on a side.}
\end{figure}

Data were reduced by bias removal, using the overscan strip at the edge
of each chip, followed by extraction and flux calibration using
standard {\sc iraf} software, distributed by the US National Optical
Astronomy Observatory (NOAO). Flux calibration was performed using a
spectrum of the standard star Hz2, obtained on a different night but
using the same instrumental setup. Wavelength calibration was done using
spectra from Hg and Cd arc lamps, and the residuals indicate that this
should be accurate to a few tenths of a nanometre except at the edges of
the blue frames. 

\section{Results}

Flux-calibrated spectra for all four candidates (A and B images in each
case) are given in Fig. 2. In each case, we identify two objects along
the slit in each spectrum, and can clearly distinguish the two spectra.
In all four systems, we identify the primary (A) object as a quasar,
with a redshift that agrees with the SDSS redshift. In two cases 
(J033248.5$-$002155 and J091750.5+290137), we clearly identify the
secondary as an M dwarf, most likely with a spectral type around 
type M5 (e.g. Bochanski et al. 2006). In the case of J034025.5$-$000820,
the identification of the object is less clear; it is hardly visible in
the blue, but the spectrum rises steeply to the red. There is a possible
identification of a break in the spectrum at around 640~nm, which if
identified with a galactic 400-nm break feature would imply that it is a
galaxy at roughly the same redshift as the quasar. In any case there is
no sign of any emission lines which might lead us to conclude that we
are dealing with a gravitational lens system.

\begin{figure}
\psfig{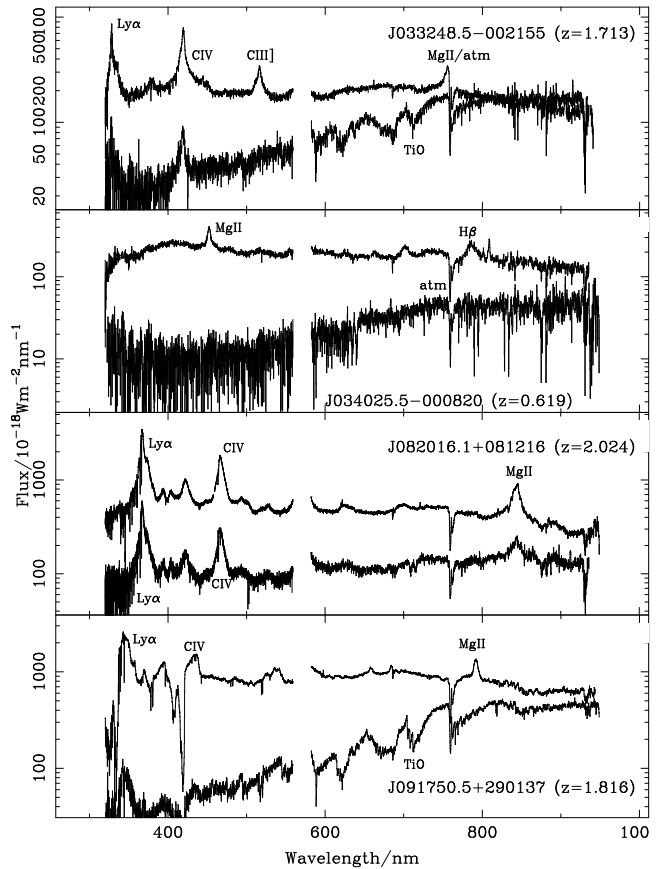}
\caption{Spectra of the four observed objects. Each panel shows the
primary object (a quasar in each case) together with the secondary. The
secondary is an M dwarf for two objects (J033248.5$-$002155 and
J091750.5+290137) and a quasar in one case (J082016.1+081216). The SDSS
redshifts are given in parentheses. Cosmic rays have been interactively
removed from the spectra, and the area affected by the dichroic cut has
been blanked. Atmospheric telluric absorption features are visible in
the spectra at 760 and 690 nm.}
\end{figure}

In the case of J082016.1+081216 (Fig. 3), we clearly see two objects with
emission lines; Ly$\alpha$, C{\sc iv} and Mg{\sc ii} are identifiable in
each spectrum, and C{\sc iii]} is hidden by the dichroic cut. Moreover, 
if we subtract a scaled version of the primary component, divided by a 
factor 6, from the secondary component, we obtain a residual which is 
redder than either spectrum individually. This is what would be 
expected from a two-image gravitational lens system, as the lensing galaxy
(G) would be expected to lie very close to the fainter image (B) of the 
lens system, with the brighter (A) image some distance away. 
The identical spectra, together with the identification of a
galactic residual in the fainter component, is convincing evidence that
this is a lens system and not, for example, a binary quasar. Unlike in
the case of ULAS~J234311.9$-$005034 (Jackson et al. 2008), there is no
evidence of any differences in the spectra which might suggest
differential reddening of the images within the lensing galaxy.
Like ULAS~J234311.9$-$005034, ULAS~J082016.1+081216 is a radio-quiet
quasar, having no radio identification at the level of 1~mJy
in the FIRST 20-cm radio survey (Becker, White \& Helfand 1995).

\begin{figure}
\psfig{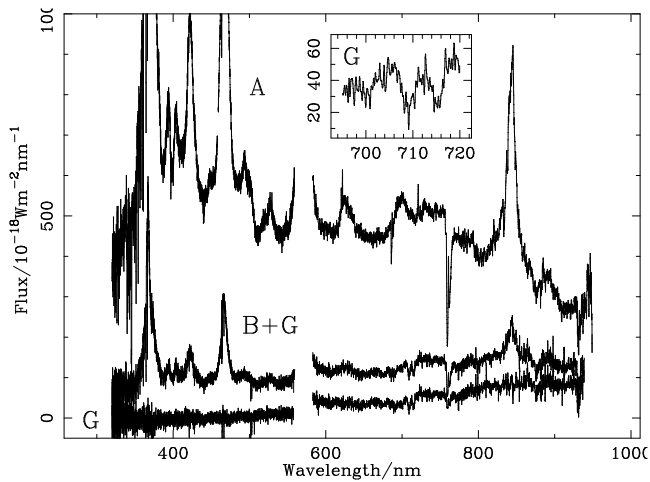}
\caption{Spectra of the J082016.1+081216 system. The figure shows the
primary component (interpreted as the brighter ``A'' image of the lens
system) and the secondary component (consisting of the ``B'' image and
the lensing galaxy G) together with the residual (G) from the subtraction
of one-sixth of the primary component from the spectrum of the secondary.
The residual is redder than either image. It contains a
possible set of absorption lines at about 710~nm (inset) which can be
identified with Ca H and K at a wavelength of 393.3, 396.7~nm in the
rest frame.
Atmospheric telluric absorption features are visible in the spectra at
760 and 690 nm.}
\end{figure}

A final indication of lensing (Fig.~4) can be derived from fitting two images to the
SDSS and UKIDSS data for J082016.1+081216. A clear trend for reduced
separation is seen between the optical and near infrared; this is
exactly as would be expected if a relatively red lensing galaxy is lying
between two blue quasar images, and close to the fainter quasar image.
The implication of Fig.~4 is that the separation of the two quasar
images is approximately 2\farcs3, and that the lensing galaxy, which
is likely to dominate the flux in the near-infrared, lies approximately
1\farcs8 from the brighter component. However, it cannot be detected
directly from the UKIDSS images alone. We can test this by fitting two
PSFs to the J-band UKIDSS image (which has the smallest pixel scale, 
0\farcs2) separated by a fixed 2\farcs27 separation implied by the blue
optical images, and allowing a third Sersic component to be located in
between them. A good fit is obtained using the {\sc galfit} software
(Peng et al. 2002), but is statistically
indistinguishable from the 2-component fit, and the residuals for the
two fits look very similar and noise-like.

\begin{figure}
\psfig{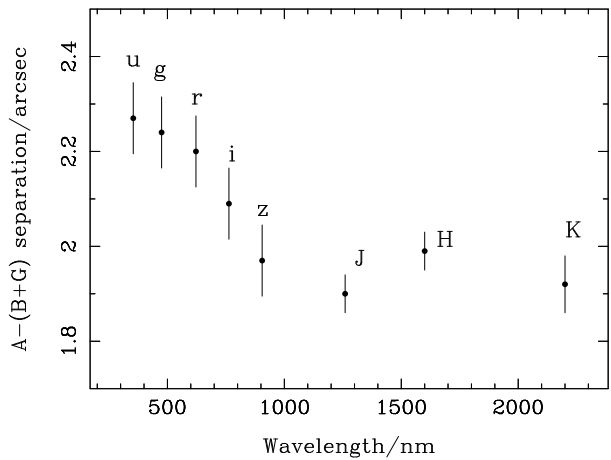}
\caption{Separation of the primary (A) and secondary (B+G) components in the 
filters $ugrizJHK$ from SDSS and UKIDSS, against wavelength} 
\end{figure}

A redshift for the galaxy can be derived if we identify the
absorption lines seen in the difference spectrum 
around 710~nm with the Ca H and K doublet at 393.3 and
396.7~nm. Fitting to these lines yields a galaxy redshift of 0.803$\pm$0.001
for each line, which, together with an Einstein radius of 1\farcs15 and
an assumption of an isothermal model, predicts a galaxy of velocity
dispersion $\sigma\simeq290$~km$\,$s$^{-1}$. 
From the Faber-Jackson relation (Faber \& Jackson 1976) as calibrated by
Rusin et al. (2003) and using the image separation together with
$z_l=0.803$, we obtain an expected magnitude of $R\simeq21.4$ for
a typical lensing galaxy. The magnitude of the galaxy implied by Fig. 3
is about 0.07 times the total magnitude of the object, or $r\simeq21.9$,
which corresponds approximately to $R=21.6$. The good agreement with the
observed $R$ is further, though circumstantial, evidence for this object
being a lens system.

If we assume an isothermal model for the galaxy, together with the
observed image flux ratio and separation, we obtain a likely time delay
of approximately 350 days, assuming $H_0=70$kms$^{-1}$Mpc$^{-1}$, 
between variations of the A and B images. The
relatively long delay results from a combination of a high flux ratio
and large separation.

\section{Discussion and conclusions}

We show that the use of the image quality together with the
depth of UKIDSS is likely to lead to discovery of lenses in a wider
region of parameter space than lenses selected using SDSS alone. 
This is because the better image quality of UKIDSS should allow the 
discovery of both smaller-separation lenses and lenses of higher flux
ratio. To illustrate this, Fig. 5 shows the image separations and flux 
ratios of lenses from the SQLS sample. For four-image lenses, the 
brightness is dominated by an almost-unresolved pair of merging images, 
with a third fainter image and a fourth, typically much fainter image. 
In this case we take the flux ratio as the brightness of the third image
divided by that of the merging pair. Fig. 5 also shows the image separation
and flux ratio distribution of lenses from the CLASS survey (Myers et al.
2003, Browne et al. 2003), which has a resolution limit of 0\farcs3 and
a flux ratio limit of 10:1, and of the two MUSCLES lenses found so far.
The lens presented here, ULAS~J082016.1+081216, has a flux ratio of 6,
higher than the limit of the SQLS main survey. In fact, of the SQLS 
optical lenses with separation $\theta<4^{\prime\prime}$, this lens 
has the highest flux ratio. Its nearest rival was found by a special 
imaging programme based on SDSS, rather than SDSS directly (Morgan et 
al. 2003).

We can extrapolate from the existing SQLS and CLASS surveys to attempt
to estimate the lens yield of MUSCLES after all followup has been done. 
Only eight of the 22 CLASS lenses lie in the part of the
separation/flux-ratio diagram accessible to the main SQLS survey.
Assuming that MUSCLES can detect lenses of up to 10:1 flux ratio, and
with separations $>0\farcs6$ (cf. the SQLS survey limit of
1$^{\prime\prime}$ for average seeing of 1\farcs4 in SDSS), this implies
a potential yield of over 50 new lenses compared to the 30 in SQLS. The
actual number may be somewhat less than this, as lenses with high flux
ratios {\em and} lower separation will be harder to detect. There will
also be a reduction because the currently planned footprint of UKIDSS 
is around 4000 square degrees, compared to around 9000 degrees in the 
SDSS spectroscopic area. It is to be hoped that extensions to UKIDSS in
the future may remedy this, however. Moreover, many of the UKIDSS 
detections of the SDSS quasars are at a level where high flux-ratio 
secondaries may be harder to find. Nevertheless, a well-selected 
lens sample approximately 2 times greater than the existing SQLS
sample has implications for studies of galaxy evolution.  For
example, the limits of Matsumoto \& Futamase (2008), based on the SQLS
sample alone together with the lens-redshift test, do not currently allow us
to rule out the hypothesis of no evolution in lens galaxy number density
or velocity dispersion. We expect that an increase 
in the statistical lens sample should allow this to be done.

\begin{figure}
\psfig{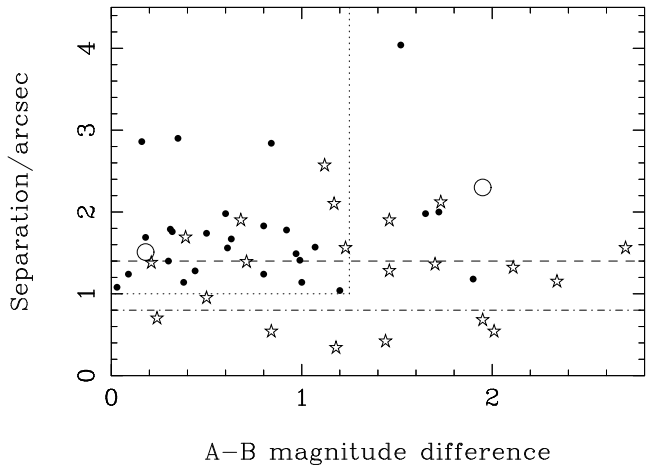}
\caption{Image separations and flux ratios from the SQLS lens sample
(Inada et al. 2003a,b, 2005, 2006a,b, 2007, 2009, 
Oguri et al. 2004, 2005, 2008a,b, Johnston et
al. 2003, Pindor et al. 2004, 2006, 
Morokuma et al. 2007, Kayo et al. 2007, Ofek et al.
2007, Morgan et al. 2003). The two MUSCLES lenses (Jackson et al. 2008
and this work) are indicated as open circles. The UKIDSS median image
quality (dot-dashed line) and SDSS (dashed line) are indicated, together
with the dynamic range and lens separation 
limit of the SDSS statistical sample (dotted line). The
primary contribution of this survey is likely to be lenses at higher
flux ratio and smaller separation. CLASS survey lenses, with a
separation limit of 0\farcs3 and flux ratio limit of about 10 (2.5
magnitudes) are indicated by stars. One CLASS lens is just outside the
plot, with a separation of 4\farcs6 and flux ratio 0.86 magnitudes.}
\end{figure}


\normalsize

\section*{Acknowledgements}

We would like to thank the Kavli Institute for Theoretical Physics
and the organizers of the KITP workshop ``Applications of Gravitational
Lensing'' for hospitality. This work began at this KITP
workshop. We thank an anonymous referee for useful comments. 
The research was supported in part by the European
Community's Sixth framework Marie Curie Research Training
Network Programme, contract no. MRTN-CT-2004-505183, by the
National Science Foundation under grant no. PHY05-51164 and
by the Department of Energy contract DE-AC02-76SF00515. This
work is based on data obtained as part of the UKIRT Infrared Deep
Sky Survey, UKIDSS (www.ukidss.org). Some of the data presented
herein were obtained at the W.M. Keck Observatory, which is operated
as a scientific partnership among the California Institute of
Technology, the University of California and the National Aeronautics
and Space Administration. The Observatory was made possible
by the generous financial support of the W.M. Keck Foundation.
The authors wish to recognize and acknowledge the very significant
cultural role and reverence that the summit of Mauna Kea has always had
within the indigenous Hawaiian community. We are most fortunate to have
the opportunity to conduct observations from this mountain. Funding for
the SDSS and SDSS-II has been provided by the Alfred P. Sloan
Foundation, the Participating Institutions, the National Science
Foundation, the US Department of Energy, the National Aeronautics and
Space Administration, the Japanese Monbukagakusho and the Max-Planck
Society, and the Higher Education Funding Council for England. The SDSS
web site is http://www.sdss.org/. The SDSS is managed by the
Astrophysical Research Consortium (ARC) for the Participating
Institutions. The Participating Institutions are the American Museum of
Natural History, Astrophysical Institute Potsdam, University of Basel,
University of Cambridge, Case Western Reserve University, The University
of Chicago, Drexel University, Fermilab, the Institute for Advanced
Study, the Japan Participation Group, The Johns Hopkins University, the
Joint Institute for Nuclear Astrophysics, the Kavli Institute for
Particle Astrophysics and Cosmology, the Korean Scientist Group, the
Chinese Academy of Sciences (LAMOST), Los Alamos National Laboratory,
the Max-Planck-Institute for Astronomy (MPIA), the Max-Planck-Institute
for Astrophysics (MPA), New Mexico State University, Ohio State
University, University of Pittsburgh, University of Portsmouth,
Princeton University, the United States Naval Observatory and the
University of Washington.

\section*{References}

\noindent Bade N., Siebert J., Lopez S., Voges W., Reimers D. 1997,  A\&A, 317, L13. 

\noindent Becker R.H., White R.L., Helfand D.J. 1995,  ApJ, 450, 559. 

\noindent Bochanski J.J., Hawley S.L., Munn J.A., Covey K.R., West A.A., Walkowicz L.M. 2006,  AAS, 20917, 214. 

\noindent Bolton A.S., Burles S., Koopmans L.V.E., Treu T., Moustakas L.A. 2006,  ApJ, 638, 703. 

\noindent Bolton A.S., Burles S., Koopmans L.V.E., Treu T., Gavazzi R., Moustakas L.A., Wayth R., Schlegel D.J. 2008,  ApJ, 682, 964. 

\noindent Browne I.W.A., Wilkinson P.N., Jackson N.J.F., Myers S.T., Fassnacht C.D., Koopmans L.V.E., Marlow D.R., Norbury M., Rusin D., Sykes C.M.,et al. 2003,  MNRAS, 341, 13. 

\noindent Capelo P.R., Natarajan P. 2007, NewJPhys 9, 445

\noindent Casali M., Adamson A., AlvesdeOliveira C., Almaini O., Burch K., Chuter T., Elliot J., Folger M., Foucaud S., Hambly N.,et al. 2007,  A\&A, 467, 777. 

\noindent Faber S.M., Jackson R.E. 1976, ApJ 204, 668

\noindent Fukugita M., Futamase T., Kasai M., Turner E.L. 1992, ApJ 393, 3

\noindent Hambly N.C., Collins R.S., Cross N.J.G., Mann R.G., Read M.A., Sutorius E.T.W., Bond I., Bryant J., Emerson J.P., Lawrence A.,et al. 2008,  MNRAS, 384, 637. 

\noindent Hewett P.C., Warren S.J., Leggett S.K., Hodgkin S.T. 2006,  MNRAS, 367, 454. 

\noindent Hodgkin S.T., Irwin M.J., Hewett P.C., Warren S.J., 2009,MNRAS, in press

\noindent Inada N., Oguri M., Pindor B., Hennawi J.F., Chiu K., Zheng W., Ichikawa S.I.,
Gregg M.D., Becker R.H., Suto Y.,et al. 2003,  Natur, 426, 810. 

\noindent Inada N., Becker R.H., Burles S., Castander F.J., Eisenstein D., Hall P.B.,
Johnston D.E., Pindor B., Richards G.T., Schechter P.L.,et al. 2003,  AJ, 126, 666. 

\noindent Inada N., Burles S., Gregg M.D., Becker R.H., Schechter P.L., Eisenstein D.J.,
Oguri M., Castander F.J., Hall P.B., Johnston D.E.,et al. 2005,  AJ, 130, 1967. 

\noindent Inada N., Oguri M., Morokuma T., Doi M., Yasuda N., Becker R.H., Richards G.T.,
Kochanek C.S., Kayo I., Konishi K.,et al. 2006,  ApJ, 653, L972. 

\noindent Inada N., Oguri M., Becker R.H., White R.L., Gregg M.D., Schechter P.L., Kawano Y.,
Kochanek C.S., Richards G.T., Schneider D.P.,et al. 2006,  AJ, 131, 1934. 

\noindent Inada N., Oguri M., Becker R.H., White R.L., Kayo I., Kochanek C.S., Hall P.B.,
Schneider D.P., York D.G., Richards G.T. 2007,  AJ, 133, 206. 

\noindent Inada N., Oguri M., Becker R.H., Shin M.S., Richards G.T., Hennawi J.F., White R.L.,
Pindor B., Strauss M.A., Kochanek C.S.,et al. 2008,  AJ, 135, 496. 

\noindent Inada N., Oguri M., Shin M., Kayo I., Strauss M.A., Morokuma
T., Schneider D.P., Becker R.H., Bahcall N.A., York D.G., 2009, astro-ph/0809.0912

\noindent Jackson N., Ofek E.O., Oguri M. 2008,  MNRAS, 387, 741. 

\noindent Johnston D.E., Richards G.T., Frieman J.A., Keeton C.R., Strauss M.A., Knapp G.R., Becker R.H., White R.L., Johnson E.T., Ma Z.,et al. 2003,  AJ, 126, 2281. 

\noindent Kayo I., Inada N., Oguri M., Hall P.B., Kochanek C.S., Richards G.T., Schneider D.P., York D.G., Pan K. 2007,  AJ, 134, 1515. 

\noindent Kochanek C.S. 1992,  ApJ, 384, 1. 

\noindent Kochanek C.S. 1996, ApJ 466, 638

\noindent Kochanek C.S. 2002, ApJ 578, 25

\noindent Koopmans L.V.E., Treu T., Bolton A.S., Burles S., Moustakas L.A. 2006,  ApJ, 649, 599. 

\noindent Lawrence A., Warren S.J., Almaini O., Edge A.C., Hambly N.C., Jameson R.F., Lucas P., Casali M., Adamson A., Dye S.,et al. 2007,  MNRAS, 379, 1599. 

\noindent Maoz D., Rix H-W. 1993, ApJ 416, 425

\noindent Matsumoto A., Futamase T. 2008,  MNRAS, 384, 843. 

\noindent Morgan N.D., Snyder J.A., Reens L.H. 2003,  AJ, 126, 2145. 

\noindent Morokuma T., Inada N., Oguri M., Ichikawa S., Kawano Y.,
Kouichi T., Kayo I., Hall P.B., Kochanek C.S., Richards G.T., York D.G.,
Schneider D.P., 2007, AJ 133, 214

\noindent Myers S.T., Jackson N.J., Browne I.W.A., deBruyn A.G., Pearson T.J., Readhead A.C.S., Wilkinson P.N., Biggs A.D., Blandford R.D., Fassnacht C.D.,et al. 2003,  MNRAS, 341, 1. 

\noindent Ofek E.O., Rix H.-W., Maoz D. 2003, MNRAS 343, 639

\noindent Ofek E.O., Oguri M., Jackson N., Inada N., Kayo I. 2007,  MNRAS, 382, 412. 

\noindent Oguri M., Inada N., Castander F.J., Gregg M.D., Becker R.H., Ichikawa S.I.,
Pindor B., Brinkmann J., Eisenstein D.J., Frieman J.A.,et al. 2004,  PASJ, 56, 399. 

\noindent Oguri M., Inada N., Keeton C.R., Pindor B., Hennawi J.F., Gregg M.D., Becker R.H.,
Chiu K., Zheng W., Ichikawa S.I.,et al. 2004,  ApJ, 605, 78. 

\noindent Oguri M., Inada N., Hennawi J.F., Richards G.T., Johnston D.E., Frieman J.A., Pindor B.,
Strauss M.A., Brunner R.J., Becker R.H.,et al. 2005,  ApJ, 622, 106. 

\noindent Oguri M., Inada N., Pindor B., Strauss M.A., Richards G.T.,
Hennawi J.F., Turner E.L., Lupton R.H., Schneider D.P., Fukugita M., Brinkmann J.
2006, AJ, 132, 999

\noindent Oguri M., Inada N., Clocchiatti A., Kayo I., Shin M.S., Hennawi J.F.,
Strauss M.A., Morokuma T., Schneider D.P., York D.G. 2008a,  AJ, 135, 520. 

\noindent Oguri M., Inada N., Blackburne J.A., Shin M., Kayo I., Strauss
M.A., Schneider D.P., York D.G., 2008b, MNRAS 391, 1973

\noindent Oke J.B., Cohen J.G., Carr M., Cromer J., Dingizian A.,
Harris F.H., Labrecque S., Lucinio R., Schaal W., Epps H., Miller J.,
1995, PASP 107, 375

\noindent Peng C.Y., Ho L.C., Impey C.D., Rix H.-W., 2002, AJ 124, 266

\noindent Pindor B., Eisenstein D.J., Inada N., Gregg M.D., Becker R.H., Brinkmann J., Burles S., Frieman J.A., Johnston D.E., Richards G.T.,et al. 2004,  AJ, 127, 1318. 

\noindent Pindor B. et al., 2006, AJ 131, 41

\noindent Refsdal S. 1964,  MNRAS, 128, 307. 

\noindent Rusin D. et al. 2003, ApJ 587, 143

\noindent Schneider D.P., Hall P.B., Richards G.T., Strauss M.A., VandenBerk D.E., Anderson S.F., Brandt W.N., Fan X., Jester S., Gray J.,et al. 2007,  AJ, 134, 102. 
\end{document}